\documentclass[reprint,amsmath,amssymb,aps]{revtex4-2} 
\usepackage{graphicx}
\usepackage{dcolumn}
\usepackage{bm}
\usepackage{floatrow}

\begin{document}
\title{Squeezing-Enhanced Two-Phase Estimation with $N$-Particle W-type States}
\author{Huan Zhang$^{1,2}$}
\author{Guofu Yin$^{3}$} 
\author{Ying Xia$^{1,2}$}
\author{Bing He$^{1,2}$} 
\author{Xiuxing Zhang$^{1}$}  
\author{Shoukang Chang$^{4}$}
\thanks{changshoukang@htu.edu.cn}
\author{Wei Ye$^{5}$}
\thanks{yeweinchu@126.com}
\affiliation{$^{{\small 1}}$\textit{School of Physics and Electrical Engineering, Weinan Normal University, Weinan 714000, China}}
\affiliation{$^{{\small 2}}$\textit{Engineering Research Center of X-ray Imaging and Detection, Universities of Shaanxi Province, Weinan Normal University, Weinan 714000, China}}
\affiliation{$^{{\small 3}}$\textit{School of Computer Science and Technology, Weinan Normal University, Weinan 714000, China}}
\affiliation{$^{{\small 4}}$\textit{School of Physics, Henan Normal University, Xinxiang 453007, China}}
\affiliation{$^{{\small 5}}$\textit{School of Information Engineering, Nanchang Hangkong University, Nanchang 330063, China}}

\begin{abstract}
We investigate the simultaneous estimation of two optical phases in a three-mode interferometer assisted by optical parametric amplification (OPA). By employing the normally ordered characteristic-function formalism, we analytically obtain all photon-number moments of the output quantum state, enabling an explicit evaluation of the quantum Fisher information matrix for multiparameter phase estimation. In the lossless scenario, we show that uniformly applied OPA significantly enhances the attainable precision beyond that of an unamplified interferometer. By analyzing the second-order correlation functions, we demonstrate that this enhancement originates from the amplification of intra-mode photon-number correlations, rather than from inter-mode correlations. We further extend our analysis to realistic interferometers with photon loss using a purification-based variational approach. Although loss degrades the achievable precision, the OPA-assisted scheme retains a clear advantage for moderate loss, indicating a degree of robustness against dissipation. Our results clarify the physical mechanism underlying OPA-enhanced multiparameter quantum metrology and provide guidelines for optimizing phase estimation protocols in realistic noisy environments.
 
\end{abstract}
\maketitle

\section{ Introduction}
Quantum metrology exploits nonclassical features of quantum systems to enhance the precision of parameter estimation beyond classical limits. By employing quantum resources such as entanglement, squeezing, and non-Gaussianity, phase sensitivities approaching the Heisenberg limit have been demonstrated in a variety of interferometric settings\cite{shimizu2025quantum,huang2025quantum,fadel2025quantum}. These advances have enabled promising applications in precision sensing, imaging, and fundamental tests of physics \cite{demille2024quantum,defienne2024advances,villegas2024optimal}. While a large body of work has focused on single-parameter estimation, realistic measurement scenarios often require the simultaneous estimation of multiple parameters encoded in a quantum system \cite{braun2018quantum,xia2023toward,valahu2025quantum}.

Extending quantum estimation from a single parameter to multiple parameters results in a considerably more complex estimation problem. In the multiparameter setting, the attainable precision is governed by a quantum Fisher information matrix (QFIM) rather than a single scalar quantity. When the generators associated with different parameters are non-commuting, a trade-off between the estimation precisions of different parameters generally arises, and the optimal strategy for one parameter may compromise the precision achievable for others \cite{hou2021zero,chen2024simultaneous,lou2025trade}. As a result, identifying probe states and interferometric configurations that enable high-precision simultaneous estimation remains a central problem in quantum metrology.

Multiarm interferometers provide a natural platform for multiparameter estimation, as multiple phases can be independently imprinted along different optical paths. In particular, multiport linear optical devices such as tritters—three-mode generalizations of beam splitters—enable coherent redistribution of optical fields and facilitate genuine multipath interference \cite{spagnolo2013three,huang2025multiphoton}. Within such architectures, path-entangled states constitute a powerful class of quantum probes. By coherently superposing distinct photon-number configurations across different paths, these states can exhibit enhanced sensitivity to phase variations due to collective interference effects \cite{hong2021quantum}. A notable example is the three-mode path-entangled Fock state, which can be regarded as a W-type generalization of the conventional N00N state to a three-arm geometry. Owing to its permutation symmetry and fixed total photon number, this state provides a balanced and unbiased resource for simultaneous estimation of multiple phases.

\begin{figure*}[tbp]
\label{Fig1} \centering \includegraphics[width=15cm]{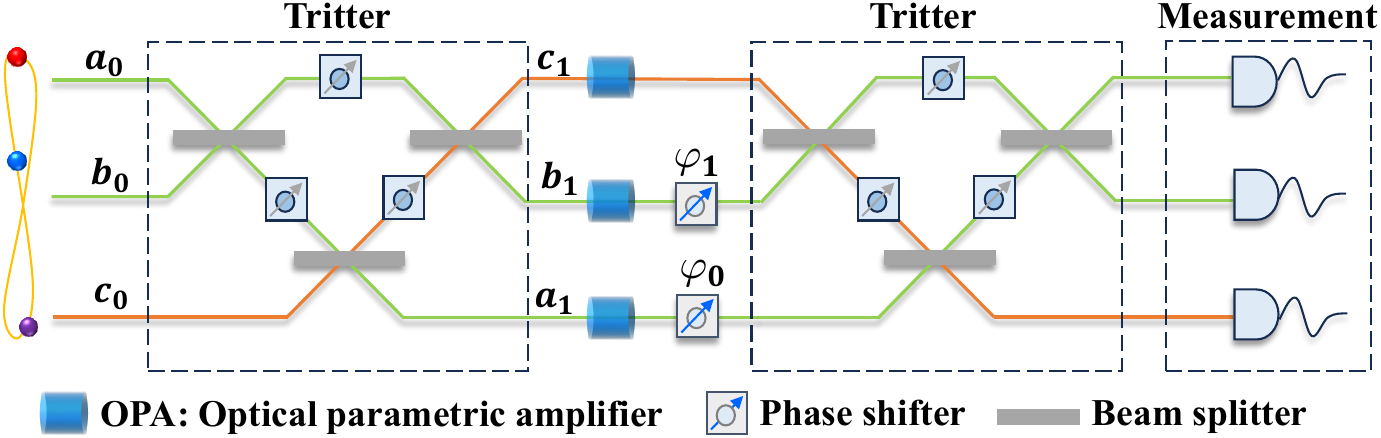}
\caption{{}(Color online) Figure 1. Schematic diagram of an OPA-assisted tritter interferometer for simultaneous two-phase estimation. A three-mode W-type state is injected into the tritter, where modes $a_0$ and $b_0$ encode the unknown phases, while mode $c_0$ serves as the reference. After the tritter, each mode undergoes single-mode optical parametric amplification followed by phase accumulation on the signal modes. The output state is finally measured to infer both phases simultaneously.}
\end{figure*}

Beyond linear interferometry, non-Gaussian resources and nonlinear optical processes offer additional avenues for enhancing metrological performance \cite{qin2023unconditional,agarwal2022quantifying,du20222}. Optical parametric amplifiers (OPAs), in particular, have been shown to effectively amplify quantum fluctuations and correlations, thereby improving phase sensitivity under suitable conditions \cite{liu2022entanglement}. The interplay between non-Gaussian probe states and parametric amplification is especially intriguing in multiparameter scenarios \cite{feng2025quantum}, where different types of quantum correlations—such as intra-mode and inter-mode photon-number correlations—may play distinct roles in determining the ultimate precision \cite{PhysRevA.96.033809}. At the same time, realistic implementations inevitably suffer from photon loss, which can severely degrade the performance of highly nonclassical states \cite{liu2022experimental,PhysRevA.108.052420,PhysRevA.103.013705}. A systematic assessment of the robustness of multiparameter quantum metrology schemes against dissipation is therefore crucial for bridging the gap between theoretical proposals and experimental feasibility.

In this work, we investigate the simultaneous estimation of two independent and identically distributed phases in a three-arm interferometric scheme driven by a three-mode path-entangled Fock-state probe. The interferometer consists of an input tritter, followed by single-mode optical parametric amplification on each arm and subsequent phase encoding. To efficiently treat the non-Gaussian nature of the probe state and its evolution under amplification, phase shifts, and photon loss, we employ the normally ordered characteristic-function formalism throughout the analysis. This approach enables us to obtain analytical expressions for photon-number moments and correlations, which directly determine the QFIM for the two-phase estimation problem.

We analyze the ultimate precision bounds in both lossless and lossy regimes. In the absence of photon loss, we show that parametric amplification can significantly enhance the estimation precision, but only when applied to the signal modes carrying phase information. In contrast, amplification applied solely to a reference mode does not lead to any improvement, highlighting the essential role of signal-mode correlations. By examining second-order photon-number correlation functions, we demonstrate that the advantage provided by OPAs originates from the enhancement of intra-mode photon-number correlations rather than inter-mode correlations. Furthermore, by comparing the path-entangled probe with separable Fock-state inputs, we explicitly identify the role of path entanglement in achieving superior multiparameter sensitivity.

The effects of photon loss are investigated using a purification-based variational approach, which allows us to derive the quantum Cramér–Rao bound (QCRB) in the presence of dissipation. We find that, although photon loss inevitably degrades the achievable precision, the estimation performance remains fully determined by photon-number variances and inter-mode covariances, which can be efficiently evaluated within the characteristic-function framework. Our results provide clear physical insight into the quantum resources responsible for enhanced simultaneous phase estimation in multiarm interferometers and offer practical guidance for the design of robust multiparameter quantum metrology schemes.

The remainder of this paper is organized as follows. In Sec. II, we describe the interferometric setup, the three-mode path-entangled input state, and the characteristic-function formalism used to model the system. Section III presents the quantum estimation theory and the derivation of the QFIM and QCRB for two-phase estimation. In Sec. IV, we analyze the estimation precision in both lossless and lossy scenarios and discuss the underlying physical mechanisms. Finally, Sec. V summarizes our results and outlines possible future directions.

\section{Theoretical model and interferometric scheme}

\subsection{Three-arm interferometer based on tritters}

We consider a three-arm optical interferometer constructed from tritters, which generalize conventional beam splitters to three optical modes and enable coherent interference among multiple paths. A tritter is a linear, lossless, three-port optical device that performs a unitary transformation on the input field operators, redistributing the optical amplitudes among three output modes with well-defined relative phases \cite{spagnolo2013three,chang2022intramode}. Such devices have attracted increasing attention in multi-mode quantum interferometry and have been experimentally realized in compact and stable platforms using femtosecond laser direct writing in integrated photonic circuits \cite{gattass2008femtosecond,osellame2011femtosecond}.

In the present scheme, the tritter is modeled as a unitary operator $U_t$ acting on the three input modes $(\hat a_0, \hat a_1, \hat a_2)$, producing the output modes $(\hat a_0', \hat a_1', \hat a_2')$. In the Heisenberg picture, the mode transformation is given by
\begin{equation}
(\hat a_0',\hat a_1',\hat a_2')^T = U_t (\hat a_0,\hat a_1,\hat a_2)^T ,
\end{equation}
where the unitary matrix $U_t$ takes the form
\begin{equation}
U_t=\frac{1}{\sqrt3}
\begin{pmatrix}
1 & e^{\frac{2i\pi}{3}} & e^{\frac{2i\pi}{3}} \\
e^{\frac{2i\pi}{3}} & 1 & e^{\frac{2i\pi}{3}} \\
e^{\frac{2i\pi}{3}} & e^{\frac{2i\pi}{3}} & 1
\end{pmatrix}.
\end{equation}
This symmetric tritter equally distributes the input quantum resources among the three output modes while introducing fixed relative phases, ensuring balanced interference among all arms of the interferometer.

Physically, such a tritter can be decomposed into a network of three balanced beam splitters and three tunable phase shifters, allowing for flexible experimental control over the mode coupling. In our scheme, the tritter plays a crucial role in coherently mixing the input modes and enabling the subsequent encoding of multiple phase parameters across different paths. By redistributing the optical fields among the three arms, it provides the necessary mode interference required for simultaneous multi-parameter phase estimation.

For clarity, throughout this work we assume ideal tritters without intrinsic loss, and possible imperfections are incorporated later through an effective photon-loss model applied to each arm. This assumption allows us to focus on the fundamental limits imposed by quantum resources and interferometric structure.

\subsection{Input state: $N$-particle W-type state}

As the quantum probe of the interferometer, we consider a three-mode path-entangled photon-number state, which can be regarded as a W-type extension of the conventional N00N state to a three-arm geometry \cite{cabello2002bell,park2025entangled}, which we refer to as a W state. The input state is given by
\begin{equation}
|\psi_{\mathrm{in}}\rangle
= \frac{1}{\sqrt{3}}
\left(
|N,0,0\rangle
+ |0,N,0\rangle
+ |0,0,N\rangle
\right),
\end{equation}
where the basis states $|N,0,0\rangle$, $|0,N,0\rangle$, and $|0,0,N\rangle$ represent configurations in which all photons are localized in one of the three modes $\hat a_0$, $\hat a_1$, or $\hat a_2$, respectively, with the total photon number conserved.

This state exhibits genuine path entanglement among the three modes and is intrinsically non-Gaussian. Owing to the coherent superposition of distinct photon-number configurations, it supports interference effects that are absent in classical or separable probe states. Moreover, the state is symmetric under permutations of the three modes, making it particularly suitable for multi-arm interferometric schemes and for the simultaneous estimation of multiple phase parameters without introducing any intrinsic bias among the arms.

When a phase shift is applied to a given path, the corresponding component of the superposition acquires a phase factor proportional to the total photon number $N$, leading to an enhanced phase response compared with unentangled Fock states. This collective phase accumulation constitutes the key resource enabling quantum-enhanced precision in multiparameter estimation.

In the following, we analyze how the W state state evolves through the tritter-based interferometer and how independent phase shifts are encoded onto different modes. To this end, in order to efficiently handle the non-Gaussian nature of the W-type state as well as the effects of amplification and photon loss, we adopt the characteristic-function formalism throughout this work.

For a general multi-mode quantum state $\hat{\rho}$, the normally ordered characteristic function is defined as \cite{scully1997quantum}
\begin{equation}
\chi^{\mathrm{N}}(\boldsymbol{\lambda}) =
\mathrm{Tr}\!\left[
\hat{\rho}\,e^{\sum_{j=0}^{2} \lambda_j \hat{a}_j^\dagger}e^{-\sum_{j=0}^{2} \lambda_j^* \hat{a}_j}\right],
\end{equation}
where $\boldsymbol{\lambda}=(\lambda_0,\lambda_1,\lambda_2)$ denotes the complex phase-space variables associated with the three modes.

Within this representation, the normally ordered characteristic function of the three-mode path-entangled Fock state in Eq.~(3) can be obtained analytically as
\begin{equation}
\chi_{N}^{\mathrm{N}}(\lambda_0,\lambda_1,\lambda_2)
= \chi_{N}^{(\mathrm{d})}(\lambda_0,\lambda_1,\lambda_2)
+ \chi_{N}^{(\mathrm{od})}(\lambda_0,\lambda_1,\lambda_2),
\end{equation}

\begin{align}
 \chi_{N}^{(\mathrm{d})}(\lambda_0,\lambda_1,\lambda_2)
&= \frac{1}{3} \sum_{i=0}^{2} L_N\big(|\lambda_i|^2\big),   \\
\chi_{N}^{(\mathrm{od})}(\lambda_0,\lambda_1,\lambda_2)
&= \frac{(-1)^N}{3\,N!} \sum_{\substack{k,l=0 \\ k \neq l}}^{2} (\lambda_k \lambda_l^*)^N .
\end{align}

where $L_N(\cdot)$ denotes the Laguerre polynomial of order $N$. The term $\chi_{N}^{(\mathrm{d})}$ arises from the diagonal Fock-state contributions of the individual paths, while the term $\chi_{N}^{(\mathrm{od})}$ originates from the off-diagonal coherence between different photon-number configurations. The latter explicitly captures the path-entangled nature of the probe state and is responsible for the nonclassical interference effects exploited in quantum-enhanced phase estimation.

This characteristic-function formulation provides a convenient and unified framework for analyzing the subsequent phase encoding, nonlinear amplification, and dissipative dynamics, which will be discussed in the following sections.

\subsection{Phase encoding and interferometric evolution}

After the preparation of the path-entangled probe state, the three optical modes are injected into the tritter-based interferometer. The first tritter coherently mixes the input modes and redistributes the photons among the three arms, enabling multi-path interference. The tritter implements a linear unitary transformation among the three optical modes, characterized by the unitary matrix $U_t$ introduced in Sec.~II.A.

Within the characteristic-function formalism, this linear optical transformation corresponds to a linear mapping of the phase-space variables,
\begin{equation}
(\lambda_{a_0'},\lambda_{a_1'},\lambda_{a_2'})^T
=
U_t^\dagger
(\lambda_{a_0},\lambda_{a_1},\lambda_{a_2})^T .
\end{equation}
Accordingly, the normally ordered characteristic function after the tritter is given by
\begin{equation}
\chi^{(1)}(\boldsymbol{\lambda})
=
\chi^{\mathrm{N}}_N
\!\left(
U_t^\dagger \boldsymbol{\lambda}
\right),
\end{equation}
where $\boldsymbol{\lambda}=(\lambda_{a_0'},\lambda_{a_1'},\lambda_{a_2'})$.

After the tritter, each optical mode undergoes an independent single-mode nonlinear amplification process described by the squeezing operator
\begin{equation}
\hat{S}_j(r)
=
\exp\!\left[
\frac{r}{2}
\left(
\hat{a}_j'^{\dagger 2}
-
\hat{a}_j'^{2}
\right)
\right],
\qquad j=0,1,2,
\end{equation}
where the squeezing parameter $r$ is assumed to be identical for all three modes. In the Heisenberg picture, this operation induces the Bogoliubov transformation
\begin{equation}
\hat{a}_j'
\rightarrow
\hat{a}_j' \cosh r + \hat{a}_j'^\dagger \sinh r .
\end{equation}

In phase space, the squeezing operation mixes each complex variable with its conjugate,
\begin{equation}
\lambda_j
\rightarrow
\lambda_j \cosh r - \lambda_j^* \sinh r ,
\end{equation}
so that the characteristic function after nonlinear amplification reads

\begin{equation}
\chi^{(2)}(\boldsymbol{\lambda})
=
\chi^{(1)}(\tilde{\lambda}_0,\tilde{\lambda}_1,\tilde{\lambda}_2),
\end{equation}
with
\begin{equation}
\tilde{\lambda}_j
=
\lambda_j \cosh r - \lambda_j^* \sinh r.
\end{equation}

Subsequently, two independent and identically distributed phase shifts $(\varphi_1,\varphi_2)$ are imprinted onto modes $\hat a_0$ and $\hat a_1$, while mode $\hat a_2$ serves as a phase reference. The phase-encoding unitary is given by
\begin{equation}
\hat{U}_{\boldsymbol{\varphi}}
=
\exp\!\left(
i \varphi_0 \hat{n}_0
+
i \varphi_1 \hat{n}_1
\right).
\end{equation}
In the characteristic-function representation, the phase shifts act as rotations of the corresponding phase-space variables,
\begin{equation}
\lambda_j \rightarrow \lambda_j e^{i\varphi_j},
\qquad j=0,1 ,
\end{equation}
with $\lambda_2$ unchanged.

Combining the above transformations, the normally ordered characteristic function after tritter mixing, nonlinear amplification, and phase encoding can be compactly written as
\begin{equation}
\chi^{(\mathrm{out})}(\boldsymbol{\lambda})
=
\chi^{\mathrm{N}}_N
\!\left(
U_t^\dagger \tilde{\boldsymbol{\lambda}}
\right),
\end{equation}
where $\tilde{\boldsymbol{\lambda}}$ denotes the set of phase-space variables transformed by the squeezing and phase-encoding operations as specified above.

Owing to the combined effects of path entanglement, nonlinear amplification, and collective phase encoding, the phase information is distributed nonlocally across the three modes. In the following sections, this output characteristic function will be used to evaluate the QFIM and the corresponding QCRB, including the effects of photon loss.

\section{Quantum estimation theory}

In this section, we present the quantum estimation framework used to assess the ultimate precision limits for the simultaneous estimation of two independent and identically distributed phase parameters. The discussion is tailored to the three-mode interferometric scheme introduced in Sec.~II and explicitly exploits the characteristic-function description of the output quantum state derived therein.

The two unknown phases $\boldsymbol{\varphi}=(\varphi_0,\varphi_1)$ are encoded onto two optical modes via unitary phase shifts generated by the photon number operators. The corresponding phase-encoding operation is described by
\begin{equation}
\hat{U}_{\boldsymbol{\varphi}}
=
\exp\!\left(
-i \varphi_0 \hat{n}_0
-i \varphi_1 \hat{n}_1
\right),
\end{equation}
where $\hat{n}_j=\hat{a}_j^\dagger \hat{a}_j$ denotes the photon number operator associated with mode $j=0,1$. For an input probe state $\hat{\rho}$, the phase-dependent output state is given by
$\hat{\rho}(\boldsymbol{\varphi})=\hat{U}_{\boldsymbol{\varphi}}\hat{\rho}\hat{U}_{\boldsymbol{\varphi}}^\dagger$.

In quantum parameter estimation theory, the ultimate precision bounds are governed by the QFIM \cite{helstrom1969quantum,liu2020quantum,PhysRevResearch.6.033315,PhysRevLett.111.070403}. For a general multi-parameter problem, the QFIM is defined through the symmetric logarithmic derivatives (SLDs) $\hat{L}_\mu$ associated with each parameter $\varphi_\mu$,
\begin{equation}
\partial_{\varphi_\mu}\hat{\rho}(\boldsymbol{\varphi})
=
\frac{1}{2}
\left(
\hat{L}_\mu \hat{\rho}(\boldsymbol{\varphi})
+
\hat{\rho}(\boldsymbol{\varphi}) \hat{L}_\mu
\right),
\end{equation}
and its elements are given by \cite{liu2020quantum}
\begin{equation}
\mathcal{F}_{\mu\nu}
=
\mathrm{Tr}
\!\left[
\hat{\rho}(\boldsymbol{\varphi})
\frac{\hat{L}_\mu \hat{L}_\nu + \hat{L}_\nu \hat{L}_\mu}{2}
\right].
\end{equation}

\begin{figure}
\label{Fig2} \centering \includegraphics[width=0.75\columnwidth]{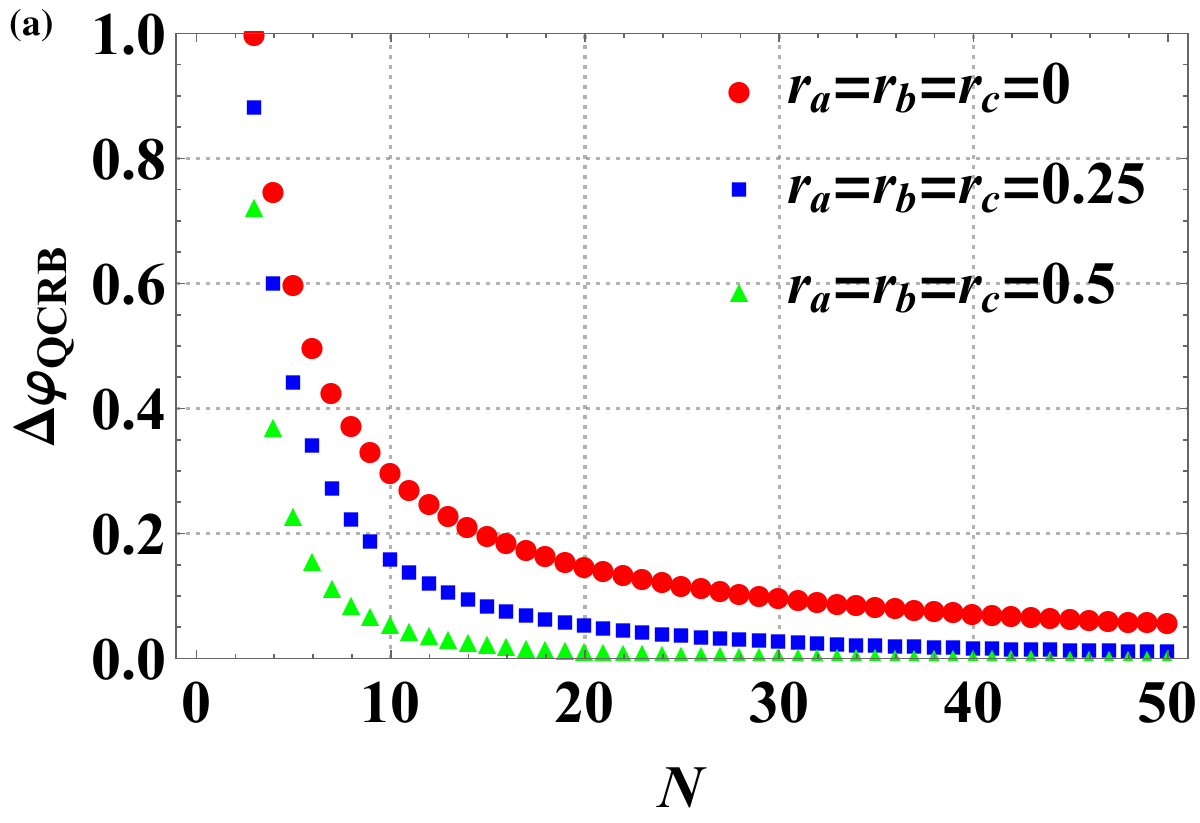}
\includegraphics[width=0.75\columnwidth]{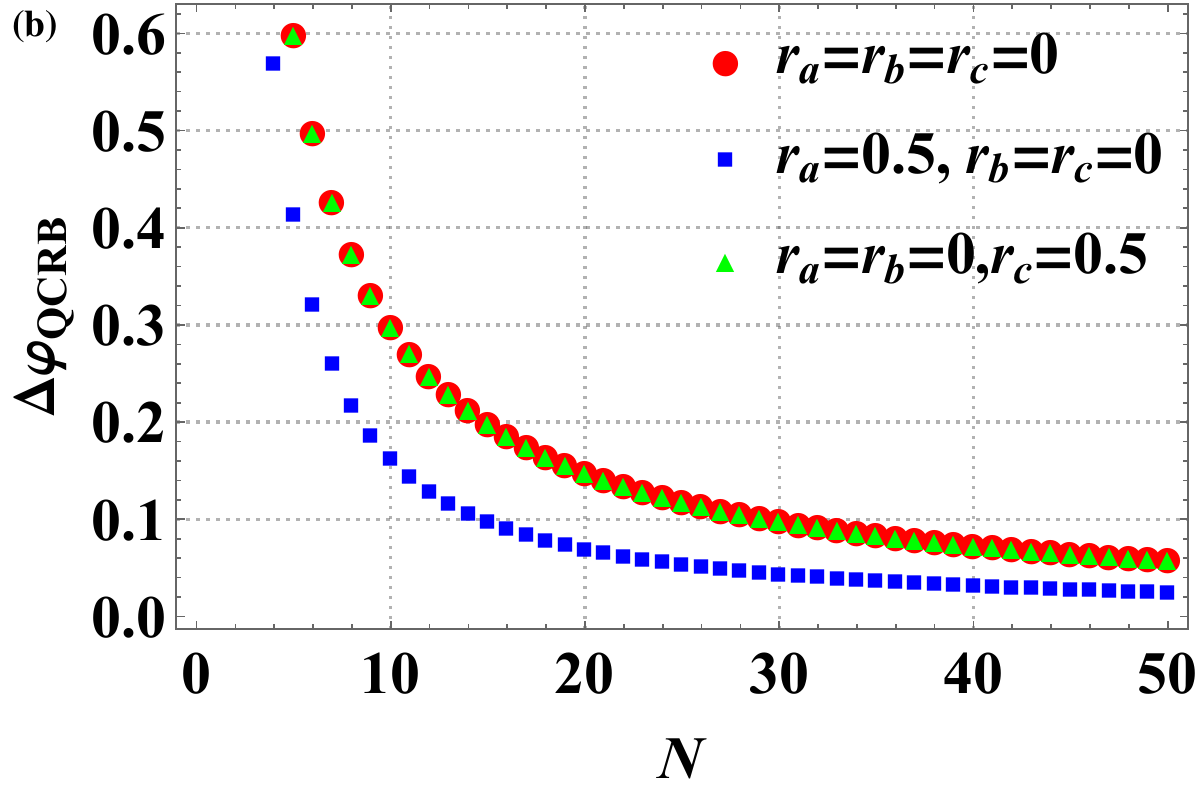}
\caption{{}(Color online)  The QCRB for simultaneous estimation of two phases as a function of the total photon number in a lossless interferometer. (a) Identical OPA gain applied to all three modes, with $r_a=r_b=r_c=0$, $0.25$, and $0.5$. (b) Mode-selective OPA configurations: $r_a=r_b=r_c=0$, $r_a=0.5,, r_b=r_c=0$, and $r_c=0.5,, r_a=r_b=0$.}
\end{figure}

Although this definition is formally general, the present problem allows for a substantial simplification. Since the phases are encoded through unitary transformations generated by the photon number operators, and since these generators commute, $[\hat{n}_a,\hat{n}_b]=0$, the SLDs can be expressed directly in terms of the generators themselves \cite{helstrom1969quantum}. As a result, the QFIM becomes independent of the phases and can be written solely in terms of the second-order moments of the photon number operators.

Specifically, for the simultaneous estimation of $\varphi_a$ and $\varphi_b$, the QFIM takes the explicit form
\begin{equation}
\mathcal{F}
=
4
\begin{pmatrix}
\left\langle \Delta^{2}\hat{n}_{0} \right\rangle
&
\operatorname{Cov}\!\left(\hat{n}_{0},\hat{n}_{1}\right)
\\[4pt]
\operatorname{Cov}\!\left(\hat{n}_{0},\hat{n}_{1}\right)
&
\left\langle \Delta^{2}\hat{n}_{1} \right\rangle
\end{pmatrix},
\end{equation}
where the variance and covariance are defined as
\begin{align}
\left\langle \Delta^{2}\hat{n}_{j} \right\rangle
&=
\langle \hat{n}_{j}^{2} \rangle
-
\langle \hat{n}_{j} \rangle^{2},
\qquad j=0,1, \\
\operatorname{Cov}\!\left(\hat{n}_{0},\hat{n}_{1}\right)
&=
\langle \hat{n}_{0}\hat{n}_{1} \rangle
-
\langle \hat{n}_{0} \rangle
\langle \hat{n}_{1} \rangle .
\end{align}

A key advantage of the characteristic-function formalism developed in Sec.~II is that all photon-number moments appearing in the QFIM can be efficiently obtained from derivatives of the normally ordered characteristic function. In particular, expectation values such as $\langle \hat{n}_j \rangle$, $\langle \hat{n}_j^2 \rangle$, and intermode correlations $\langle \hat{n}_0 \hat{n}_1 \rangle$ are directly related to derivatives of the characteristic function evaluated at the origin of phase space.

More explicitly, the photon-number moments required to construct variances and covariances can be systematically derived from the characteristic function of the output state, $\chi^{(\mathrm{out})}(\boldsymbol{\lambda})$, by applying suitable differential operators. For instance, the joint photon-number moments of modes $a_0$ and $a_1$ can be expressed as

\begin{equation}
\left\langle 
\hat{a}_0^{\dagger m} \hat{a}_0^{m}
\hat{a}_1^{\dagger n} \hat{a}_1^{n}
\right\rangle
=
\partial_{m,n}
\chi^{(\mathrm{out})}(\boldsymbol{\lambda}),
\label{eq:moment_from_CF}
\end{equation}
where the differential operator is defined as
\begin{equation}
\partial_{m,n}
=
\left.
\frac{
\partial^{2(m+n)}
}{
\partial \lambda_{a_0}^{m}
\partial(-\lambda_{a_0}^{*})^{m}
\partial \lambda_{a_1}^{n}
\partial(-\lambda_{a_1}^{*})^{n}
}
\right|_{\lambda_{a_0}=\lambda_{a_1}=\lambda_{a_2}=0}.
\end{equation}

As special cases of Eq.~(\ref{eq:moment_from_CF}), the first- and second-order photon-number moments, as well as the intermode correlations, can be readily obtained by choosing $(m,n)=(1,0)$, $(2,0)$, and $(1,1)$, respectively. These moments directly yield the mean photon numbers, variances, and covariances entering the QFIM, thereby closing the link between the characteristic-function description of the interferometric evolution and the quantum estimation bounds. This derivative-based approach avoids explicit reconstruction of the density matrix and remains computationally efficient even in the presence of non-Gaussian states, nonlinear transformations, and photon loss.

Once the QFIM is known, the ultimate lower bound of any unbiased estimator is given by the QCRB \cite{PhysRevA.95.032321,liu2016quantum},
\begin{equation}
\Delta\boldsymbol{\varphi}\ge\Delta\boldsymbol{\varphi}_\text{QCRB}=\mathrm{Tr}\left(\mathcal{F}^{-1}\right),
\end{equation}
 The diagonal elements of $\mathcal{F}^{-1}$ provide the minimum achievable variances for the individual phase estimates, while the off-diagonal elements quantify the estimation correlations between the two phase parameters. In the following section, these bounds are used to quantitatively assess the metrological performance of the proposed interferometric scheme under various physical conditions. This derivative-based approach avoids explicit reconstruction of the density matrix and remains computationally efficient even in the presence of non-Gaussian states, nonlinear transformations, and photon loss.

\section{Results and discussion}

In this section, we present and discuss the metrological performance of the proposed three-mode interferometric scheme for the simultaneous estimation of two independent phases. Based on the QFIM derived in Sec.~III and evaluated using the characteristic-function approach introduced in Sec.~II, we analyze the achievable precision bounds under various physical conditions. Particular attention is paid to the roles of path entanglement, nonlinear amplification, and photon loss in determining the estimation precision.

\subsection{Phase estimation without photon loss}

\begin{figure}
\label{Fig3} \centering \includegraphics[width=0.75\columnwidth]{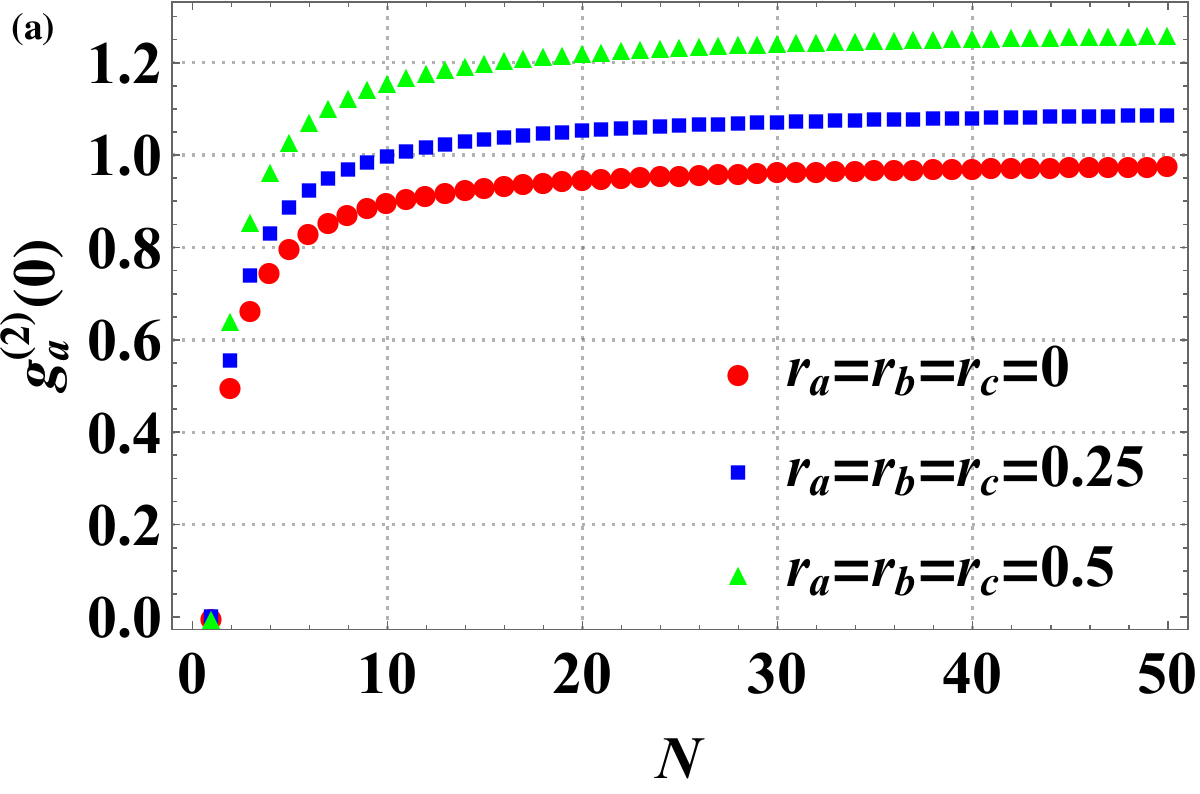}
\includegraphics[width=0.75\columnwidth]{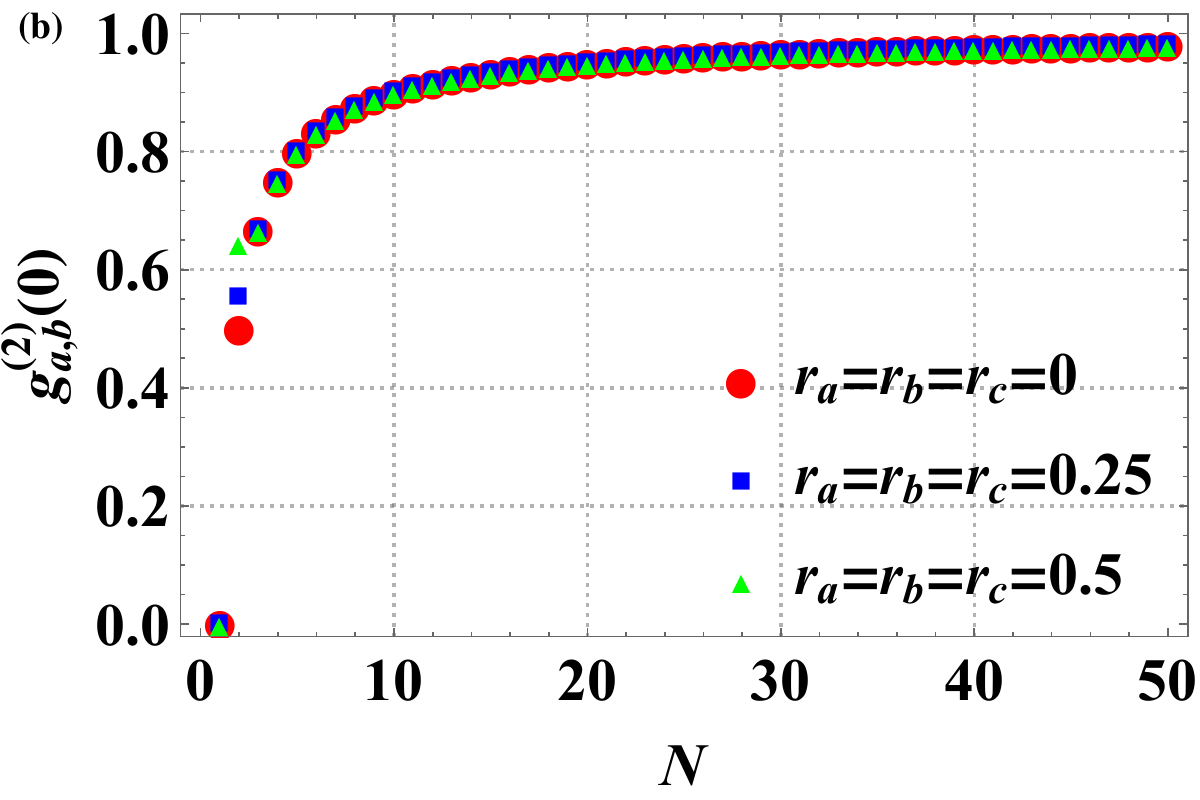}
\caption{{}(Color online)  Second-order photon-number correlations of the output state in a lossless interferometer. (a) Intra-mode second-order correlation function $g^{(2)}a$ versus the total photon number for identical OPA gain $r_a=r_b=r_c=0$, $0.25$, and $0.5$. (b) Inter-mode second-order correlation function $g^{(2)}{a,b}$ for the same gain configurations as in panel (a).}
\end{figure}

We first analyze the ideal scenario in which photon loss is absent, such that the interferometric evolution remains unitary and the metrological performance is solely determined by the properties of the output quantum state. In this regime, the QFIM derived in Sec.~III provides the ultimate precision limits for the simultaneous estimation of the two phases.

As shown in Fig.~2(a), when all three modes are uniformly amplified before the phase encoding stage, the QCRB decreases significantly with increasing OPA gain. For a fixed total photon number $N$, stronger nonlinear amplification leads to a systematically lower estimation error, indicating a clear metrological advantage brought by the OPA. This improvement persists over a wide range of photon numbers, demonstrating that the enhancement is not restricted to the few-photon regime.

Fig.~2(b) further reveals that this enhancement is highly mode selective. When the OPA is applied only to one of the phase-encoding signal modes, a noticeable improvement of the QCRB is still observed. In contrast, when the amplification is restricted solely to the reference mode while the signal modes remain unamplified, the estimation precision essentially coincides with the unamplified case. This behavior shows that amplifying a mode that does not directly carry phase information does not contribute to improving the sensitivity. Therefore, the observed enhancement cannot be attributed to a trivial increase of optical energy or global squeezing, but rather to the action of the OPA on the phase-encoding modes themselves.

To elucidate the physical mechanism underlying these observations, we analyze the second-order photon correlation functions. Fig.~3(a) displays the intra-mode second-order correlation function $g^{(2)}_a(0)$ as a function of the photon number for different uniform amplification strengths. It is evident that increasing the OPA gain leads to a pronounced enhancement of intra-mode photon-number correlations, with $g^{(2)}_a(0)$ growing monotonically and saturating at higher photon numbers. Importantly, the qualitative dependence of $g^{(2)}_a(0)$ on both $N$ and the gain closely mirrors the behavior of the QCRB shown in Fig.~2(a).

By contrast, Fig.~3(b) shows the inter-mode second-order correlation function $g^{(2)}_{a,b}(0)$ for the same set of parameters. Although weak inter-mode correlations are present, their dependence on the OPA gain is much less pronounced, and all curves tend to converge rapidly with increasing photon number. Notably, no clear correspondence is observed between the behavior of $g^{(2)}_{a,b}(0)$ and the enhancement of the QCRB. This indicates that correlations between different optical paths are not the dominant resource responsible for the improved estimation precision.

Taken together, Figs.~2 and 3 provide a consistent physical picture. The nonlinear amplification enhances phase sensitivity primarily by increasing photon-number correlations within each phase-encoding mode, as quantified by the intra-mode correlation function $g^{(2)}_a(0)$. These enhanced fluctuations directly enlarge the photon-number variances entering the QFIM. In contrast, inter-mode correlations or amplification of non–phase-encoding reference modes do not play a significant role. The metrological advantage therefore stems from the amplification-induced strengthening of local photon correlations in the signal modes, rather than from global entanglement redistribution or energy injection alone.

These results demonstrate that the metrological advantage introduced by the OPA-assisted tritter interferometer originates predominantly from enhanced intra-mode photon-number correlations, as quantified by the second-order correlation function $g^{(2)}_a$, rather than from correlations between different modes. The nonlinear amplification reshapes the photon-number statistics within each mode in a way that directly increases the photon-number variances entering the QFIM, thereby leading to a reduced QCRB.

\begin{figure}
\label{Fig4} \centering \includegraphics[width=0.75\columnwidth]{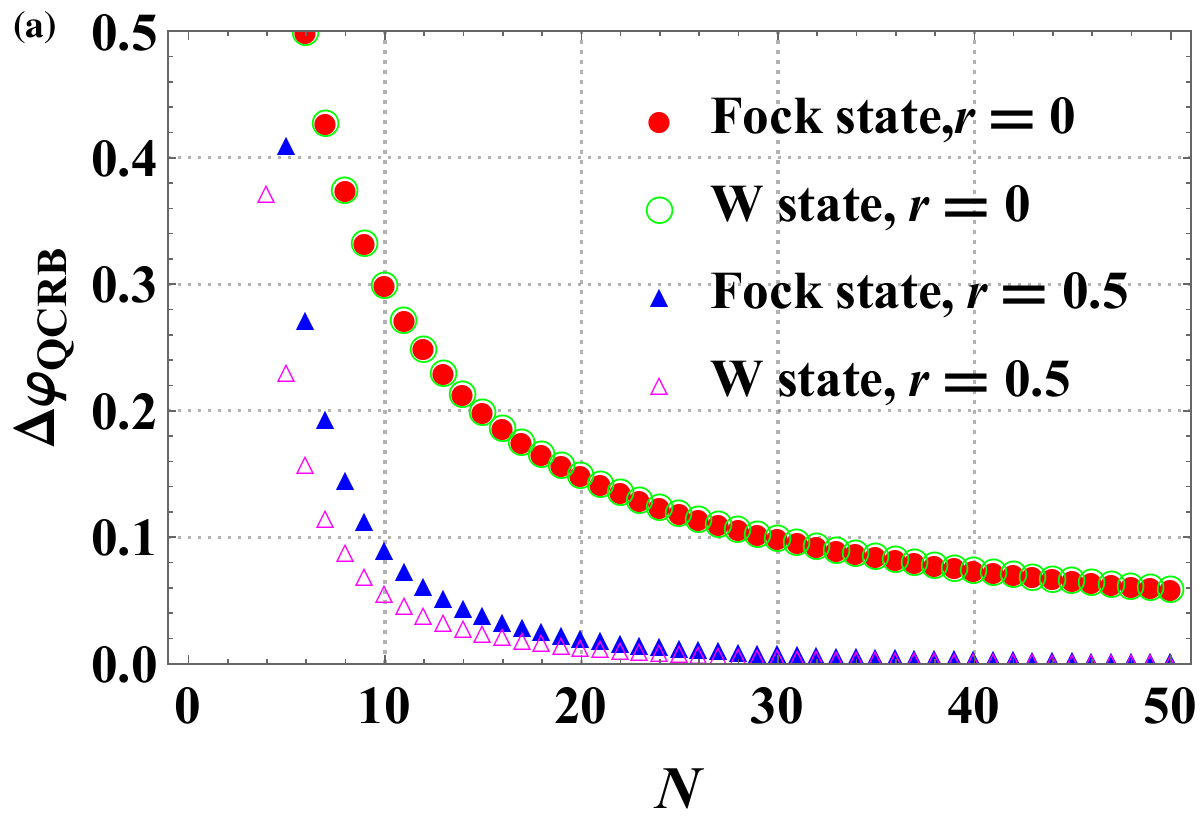}
\includegraphics[width=0.75\columnwidth]{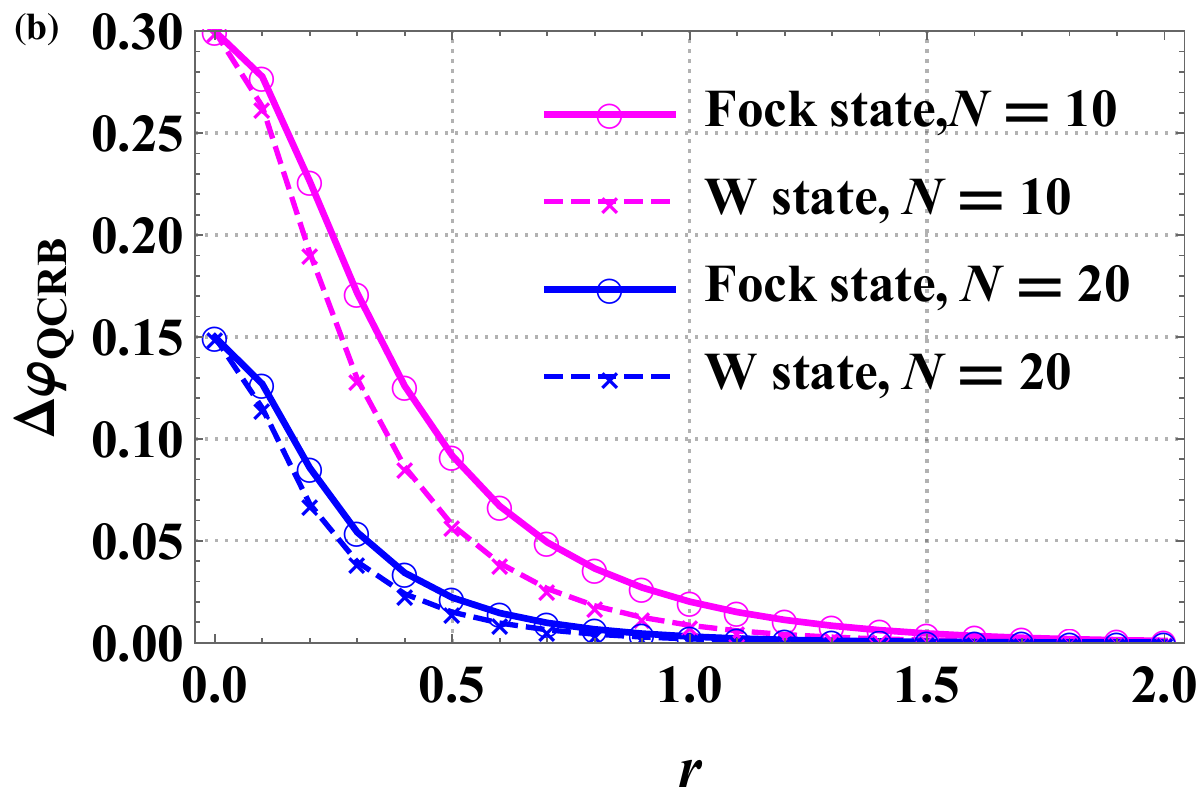}
\caption{{}(Color online)  Comparison of phase-estimation performance between a three-mode W state and a separable Fock state in a lossless interferometer. The QCRB is shown as a function of the total photon number for both input states under identical interferometric and OPA conditions.}
\end{figure}

To further clarify the role of quantum correlations in the input probe state, we compare the performance of the three-mode path-entangled W state with that of a separable Fock state, as shown in Fig.~4. The separable Fock probe is constructed by injecting all photons into a single mode, without any path entanglement.

Figure~4(a) illustrates the dependence of the QCRB on the photon number for both probe states, with and without nonlinear amplification. In the absence of OPA ($r=0$), the W state consistently outperforms the separable Fock state, highlighting the metrological advantage stemming from path entanglement. When OPA is introduced ($r=0.5$), the precision enhancement becomes significantly more pronounced for the W state, whereas the improvement for the separable Fock state remains comparatively modest.

This contrast demonstrates that nonlinear amplification does not uniformly enhance all probe states. Instead, the OPA is far more effective when acting on a quantum probe states that already possesses nonclassical correlations. In the case of the W state, the coherent superposition of different photon-number configurations allows the amplification process to strengthen photon-number fluctuations in each phase-encoding mode, which directly contributes to the quantum Fisher information. By comparison, the separable Fock state lacks such coherent path superpositions, limiting the ability of OPA to generate useful correlations for phase estimation.

Fig.~4(b) further confirms this interpretation by showing the QCRB as a function of the amplification strength for fixed photon numbers. As the gain increases, the estimation precision improves much more rapidly for the W state than for the separable Fock state. This indicates that the interplay between path entanglement and nonlinear amplification is essential for achieving substantial metrological advantages. OPA alone, without an appropriate quantum-correlated input, is insufficient to realize the same level of enhancement.

Overall, Fig.~4 demonstrates that the optimal performance observed in Fig.~2 arises from a cooperative effect between the path-entangled structure of the W state and the nonlinear amplification provided by the OPA. The W state provides the necessary nonclassical correlations, while the OPA selectively amplifies the photon-number fluctuations within the phase-encoding modes, leading to a substantial reduction of the QCRB.

\subsection{Phase estimation with photon loss}

In realistic interferometric implementations, photon loss is unavoidable and generally degrades the attainable precision of phase estimation. In the presence of optical dissipation, the phase-encoding process becomes intrinsically nonunitary, and the output probe state is no longer pure \cite{PhysRevA.110.052615,PhysRevLett.133.190801,PhysRevX.12.011039}. Consequently, the QFIM derived for ideal, lossless interferometers cannot be directly applied.

To rigorously quantify the ultimate precision bound under photon loss, we adopt a purification-based variational approach \cite{PhysRevLett.109.190404}. Within this framework, photon loss is modeled by coupling each optical mode of the system to an independent environmental mode via a fictitious beam splitter \cite{PhysRevA.110.012460}. Although the reduced dynamics of the system alone is nonunitary, the joint system--environment evolution remains unitary, enabling a systematic evaluation of the QCRB.

The purified state after phase encoding can be written as
\begin{equation}
|\Psi(\boldsymbol{\varphi})\rangle_{SE}
=
\hat{U}_{SE}(\boldsymbol{\varphi})
|\psi_{\mathrm{in}}\rangle_S
\otimes
|0\rangle_E ,
\end{equation}
where $\boldsymbol{\varphi}=(\varphi_0,\varphi_1)$ denotes the two independent phases to be estimated. The global unitary operator factorizes as
\begin{equation}
\hat{U}_{SE}(\boldsymbol{\varphi})
=
\hat{U}^{(a_0)}_{SE}(\varphi_0)
\otimes
\hat{U}^{(a_1)}_{SE}(\varphi_1)
\otimes
\hat{U}^{(a_2)}_{SE},
\end{equation}
with the reference mode $a_2$ carrying no phase information. Tracing over the environmental degrees of freedom gives rise to a Kraus representation for the reduced system state. While the joint evolution is governed by the unitary operator $\hat{U}_{SE}$, the effective nonunitary dynamics of the system alone is fully characterized by a set of Kraus operators $\{\hat{K}_{\ell}\}$ \cite{escher2011general}, yielding
\begin{equation}
\rho(\boldsymbol{\varphi})
=
\sum_{\ell}
\hat{K}_{\ell}(\boldsymbol{\varphi})
\rho_{\mathrm{in}}
\hat{K}_{\ell}^{\dagger}(\boldsymbol{\varphi}),
\end{equation}
where $\ell=(\ell_{a_0},\ell_{a_1},\ell_{a_2})$ labels the photon-loss events in each mode and
\begin{equation}
\hat{K}_{\ell}(\boldsymbol{\varphi})
=
\hat{K}^{({a_0})}_{\ell_{a_0}}(\varphi_0)
\otimes
\hat{K}^{({a_1})}_{\ell_{a_1}}(\varphi_1)
\otimes
\hat{K}^{({a_2})}_{\ell_{a_2}}.
\end{equation}

Within this purification framework, the QCRB for simultaneous estimation of the two phases is given by
\begin{equation}
|\Delta \boldsymbol{\varphi}|_{\mathrm{QCRB}}^{2}
=
\max_{\{\hat{K}_{\ell}(\boldsymbol{\varphi})\}}
\operatorname{Tr}
\left[
\mathcal{C}^{-1}
\bigl(\boldsymbol{\varphi},\hat{K}_{\ell}\bigr)
\right],
\label{eq:QCRB_loss_variational}
\end{equation}
where $\mathcal{C}$ denotes the QFIM associated with a given Kraus representation \cite{PhysRevA.106.062409}.

For symmetric photon loss, characterized by identical transmissivity $\eta$ and identical loss-position parameter $\sigma$ for all interferometric paths, the elements of the QFIM can be cast into the compact form
\begin{equation}
\mathcal{C}_{jk}
=
4\left(
\langle \hat{\mathcal{A}}_{jk} \rangle
-
\langle \hat{\mathcal{B}}_{j} \rangle
\langle \hat{\mathcal{B}}_{k} \rangle
\right),
\qquad j,k\in\{0,1\},
\end{equation}
with
\begin{equation}
\hat{\mathcal{B}}_{j}
=
\mu_j\,\hat{n}_j ,
\qquad
\hat{\mathcal{A}}_{jj}
=
\mu_j^{2}\hat{n}_j^{2}
+
\delta_j \hat{n}_j ,
\end{equation}
and $\hat{\mathcal{A}}_{jk}=\hat{\mathcal{B}}_{j}\hat{\mathcal{B}}_{k}$ for $j\neq k$.
The coefficients originate from the Kraus representation of photon loss and read
\begin{equation}
\mu_j = 1-\sqrt{\frac{\delta_j(1-\eta_j)}{\eta_j}},
\qquad
\delta_j = \eta_j(1-\eta_j)(1+\sigma_j)^2 .
\end{equation}

In the following, we focus on a symmetric loss scenario with $\eta_j=\eta$ and $\sigma_j=\sigma$ for all modes $j\in\{{a_0},{a_1},{a_2}\}$.
Under this assumption, the total estimation error obtained by explicitly inverting the $2\times2$ QFIM is given by
\begin{equation}
\operatorname{Tr}\!\left[\mathcal{C}^{-1}\right]
=
\frac{1}{2\Xi^{2}}
\frac{
(\Delta^{2}\hat{n}_{0})\,\Omega
}{
(\Delta^{2}\hat{n}_{0})^{2} \Omega^{2}
-
\left[\operatorname{Cov}(\hat{n}_{0},\hat{n}_{1})\right]^{2}
}.
\label{eq:QCRB_loss_final}
\end{equation}

Here, the loss-dependent coefficients are defined as
\begin{align}
\Xi &= 1-(1-\eta)(1+\sigma), \\
\Omega &=
1+
\frac{\eta(1-\eta)(1+\sigma)^{2}}{\Xi^{2}}
\frac{\langle \hat{n}_{0} \rangle}{\Delta^{2}\hat{n}_{0}} .
\end{align}

\begin{figure}
\label{Fig5} \centering \includegraphics[width=0.75\columnwidth]{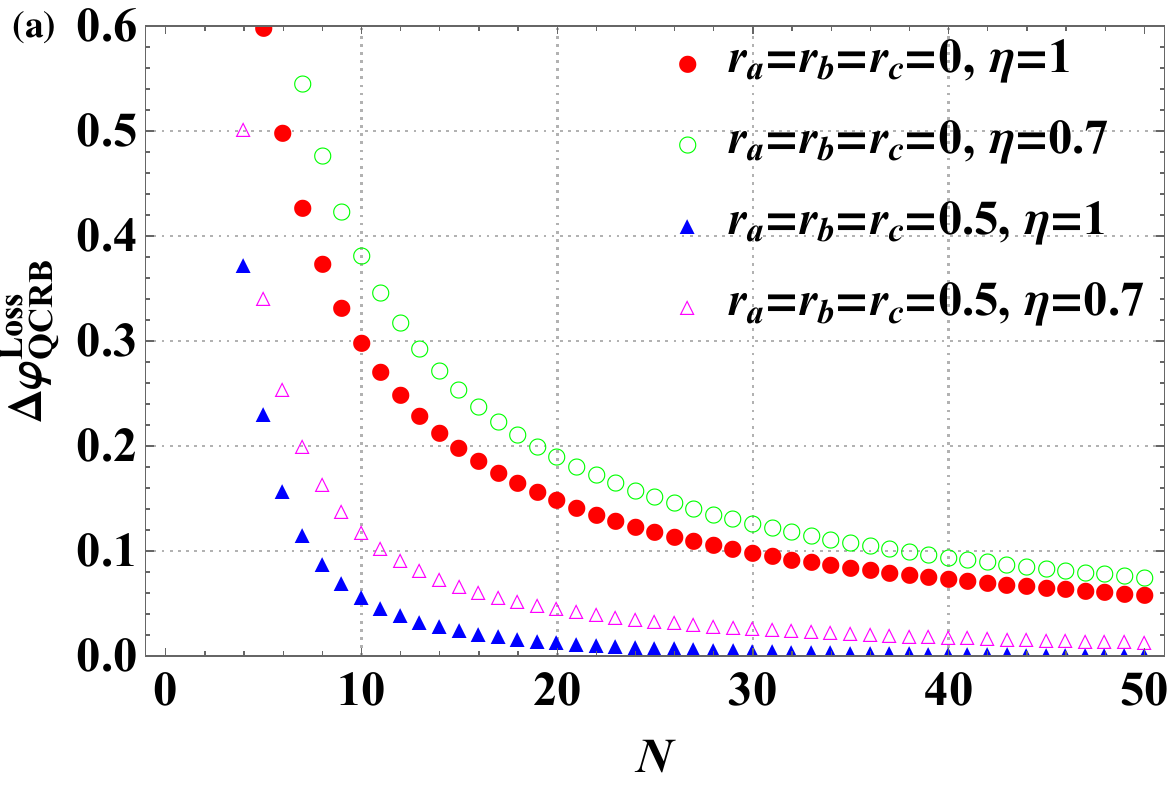}
\includegraphics[width=0.75\columnwidth]{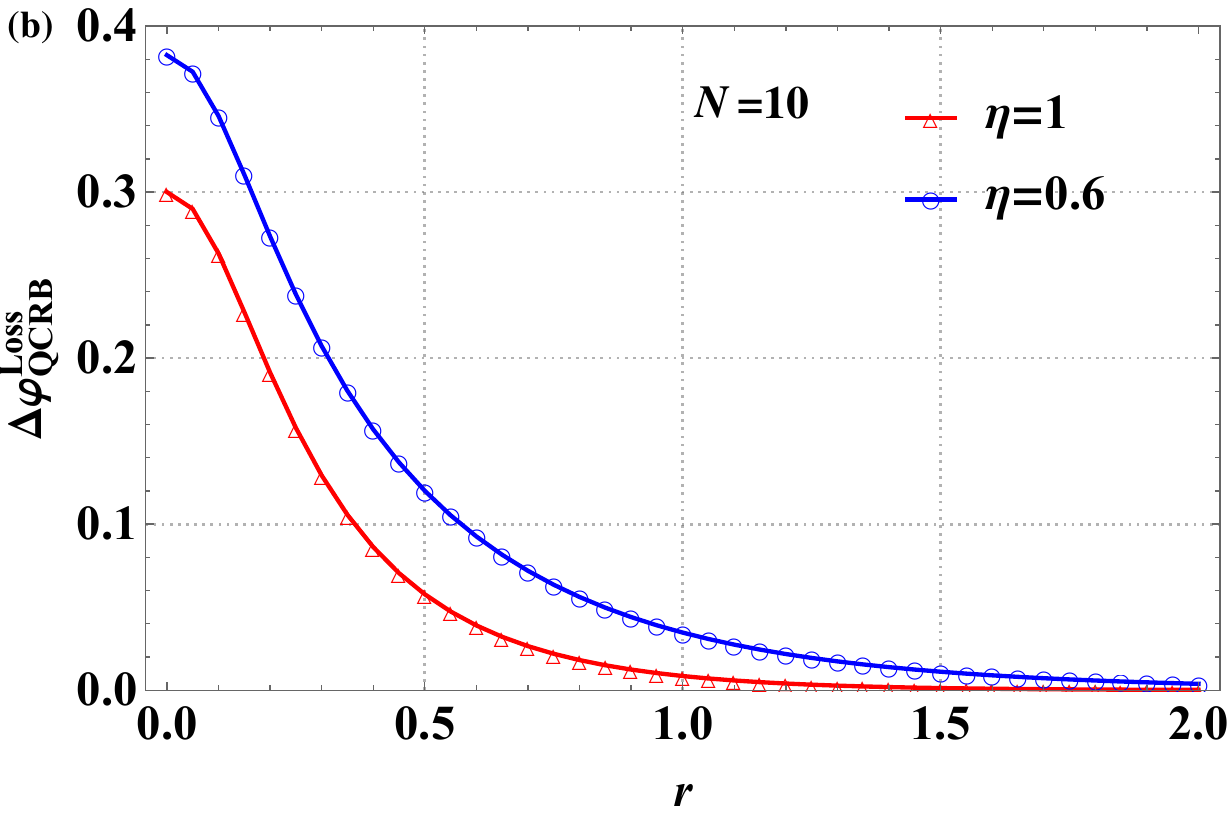}
\caption{{}(Color online) Effects of photon loss on simultaneous two-phase estimation.
(a) QCRB as a function of the total photon number \(N\) for different OPA gains and photon transmissivities, with \(r_a=r_b=r_c=0,\,\eta=1\); \(r_a=r_b=r_c=0,\,\eta=0.7\); \(r_a=r_b=r_c=0.5,\,\eta=1\); and \(r_a=r_b=r_c=0.5,\,\eta=0.7\).
(b) QCRB as a function of the OPA gain \(r\) for a fixed photon number \(N=10\), shown for two different transmissivities \(\eta=1\) and \(\eta=0.6\).}
\end{figure}

Equ.~(\ref{eq:QCRB_loss_final}) explicitly shows that, even in the presence of photon loss, the attainable estimation precision is fully determined by the photon-number variance and the inter-mode photon-number covariance of the probe state. Importantly, these quantities can be efficiently evaluated using the normally ordered characteristic function developed in Sec.~II, enabling a direct and systematic analysis of loss-induced degradation.

With the general expression of the QCRB under photon loss derived above, we now analyze how optical dissipation affects the achievable precision of simultaneous two-phase estimation. The numerical results are summarized in Fig.~5, where the total estimation error $|\Delta \boldsymbol{\varphi}|^2_{\rm QCRB}$ is evaluated for different photon numbers, OPA gains, and loss strengths.

Figure~5(a) shows the dependence of the QCRB on the total mean photon number $N$ for both lossless ($\eta=1$) and lossy ($\eta<1$) interferometers. In the absence of loss, increasing the OPA gain uniformly applied to all modes leads to a clear improvement in phase sensitivity, consistent with the lossless analysis discussed in Sec.~IV.A. When photon loss is introduced, the overall precision is inevitably degraded, and the QCRB curves are shifted upward. Nevertheless, a noticeable metrological advantage induced by OPA remains observable even at moderate loss levels. In particular, for $\eta=0.7$, the amplified probe still outperforms the unamplified one across the entire photon-number range considered, indicating that the enhancement provided by OPA is not completely washed out by dissipation.

Importantly, Fig.~5(a) also reveals that the relative improvement due to OPA becomes less pronounced as $N$ increases in the presence of loss. This behavior can be directly understood from Eq.~(\ref{eq:QCRB_loss_final}), where photon loss enters through the coefficients governing the effective photon-number variance. While OPA increases the intrinsic photon-number fluctuations of each mode, loss suppresses these fluctuations by randomly removing photons, thereby reducing the metrologically relevant intra-mode correlations that underpin the enhancement mechanism. As a result, although parametric amplification improves robustness against loss compared with the unamplified case, it cannot fully restore the lossless scaling at large photon numbers.

Figure~5(b) further illustrates the interplay between OPA gain and photon loss by fixing the photon number and varying the amplification parameter $r$. For an ideal interferometer, increasing $r$ monotonically improves the estimation precision, reflecting the continuous growth of photon-number variance induced by OPA. In contrast, under photon loss the performance improvement saturates and eventually becomes marginal at large $r$. This saturation originates from the fact that loss imposes an effective upper bound on the usable photon-number fluctuations, beyond which additional amplification mainly amplifies photons that are subsequently lost to the environment, contributing little to phase sensitivity.

Taken together, these results demonstrate that photon loss quantitatively limits the achievable precision but does not qualitatively alter the underlying enhancement mechanism. Even in the lossy regime, the QCRB is still governed by the photon-number variance and inter-mode covariance of the input state, as predicted by the analytical expression in Eq.~(\ref{eq:QCRB_loss_final}). In particular, the residual advantage provided by OPA stems from its ability to enhance intra-mode photon-number correlations, which are more robust against moderate loss than inter-mode correlations. This confirms that the physical origin of the precision enhancement identified in the lossless case persists in realistic interferometric environments.

\section{Conclusion}
In this work, we have studied simultaneous two-phase estimation in a three-mode interferometric scheme assisted by optical parametric amplification. Using the characteristic-function formalism, we derived compact expressions for photon-number moments after the tritter, OPA, and phase-encoding stages, which allowed us to obtain the QFIM in a transparent and analytically tractable form.

In the absence of photon loss, we demonstrated that OPA can substantially improve the estimation precision, as quantified by the QCRB. By comparing the behavior of the QCRB with that of second-order correlation functions, we identified the physical origin of this enhancement as the increase of intra-mode photon-number correlations induced by parametric amplification. Notably, we found that amplification applied solely to the reference mode does not improve phase sensitivity, highlighting the essential role of signal-mode correlations. Moreover, our analysis shows that inter-mode correlations play a negligible role in determining the achievable precision in this scheme.

We further investigated the impact of photon loss on simultaneous phase estimation by employing a purification-based variational method. While photon loss inevitably degrades the estimation precision, the OPA-assisted interferometer retains a noticeable advantage over the unamplified case under moderate loss. The resulting precision bound remains fully determined by the photon-number variance and inter-mode covariance of the input state, which can be efficiently evaluated within the characteristic-function framework.

Our results provide a unified understanding of OPA-enhanced multiparameter phase estimation, both in ideal and lossy interferometers. They clarify the role of quantum correlations in determining the ultimate precision limits and offer practical insights into the design of robust quantum metrological protocols in realistic experimental conditions.

\section{Acknowledgments}
	This work was supported by the Shaanxi Fundamental Science Research Project for Mathematics and Physics (Grant No.23JSY023),  the Shaanxi Provincial Key Research and Development Program (Project No. 2025CY-YBXM-068), the Shaanxi Provincial Department of Education Key Scientific Research Project (Project No. 24JR060), the Jiangxi Provincial Natural Science Foundation (Grant No. 20232BAB211032, 20252BAC240169), and the Scientific Research Startup Foundation (Grant No. EA202204230).

\end{document}